\documentclass[
    ,final            % use final for the camera ready runs
  ]
  {aipproc}

\layoutstyle{6x9}

\begin{document}

\title{The ARGO$-$YBJ sensitivity to GRBs}

\author{T. Di Girolamo}{
  address={INFN, Napoli, Italy}
}

\author{G. Di Sciascio}{
  address={INFN, Napoli, Italy}
}

\author{S. Vernetto}{
  address={IFSI-CNR and INFN, Torino, Italy}
} 

%\author{the ARGO$-$YBJ Collaboration}{}

\begin{abstract}
ARGO$-$YBJ is a ``full coverage'' air shower detector under construction at
the YangBaJing Laboratory (4300 $m$ a.s.l., Tibet, P.R. of China). Its main 
goals are $\gamma$-ray astronomy and cosmic ray studies. In this paper we
present the capabilities of ARGO$-$YBJ in detecting the emission from
Gamma Ray Bursts (GRBs) at energies $E>10\;GeV$.
\end{abstract}

\maketitle

%%%%%%%%%%%%%%%%%%%%%%%%%%%%%%%%%%%%%%%%%%%%
%% MAINMATTER
%%%%%%%%%%%%%%%%%%%%%%%%%%%%%%%%%%%%%%%%%%%%

\section{The experiment}

\begin{figure}
\begin{minipage}[t]{.6\columnwidth}
  \includegraphics[height=.3\textheight]{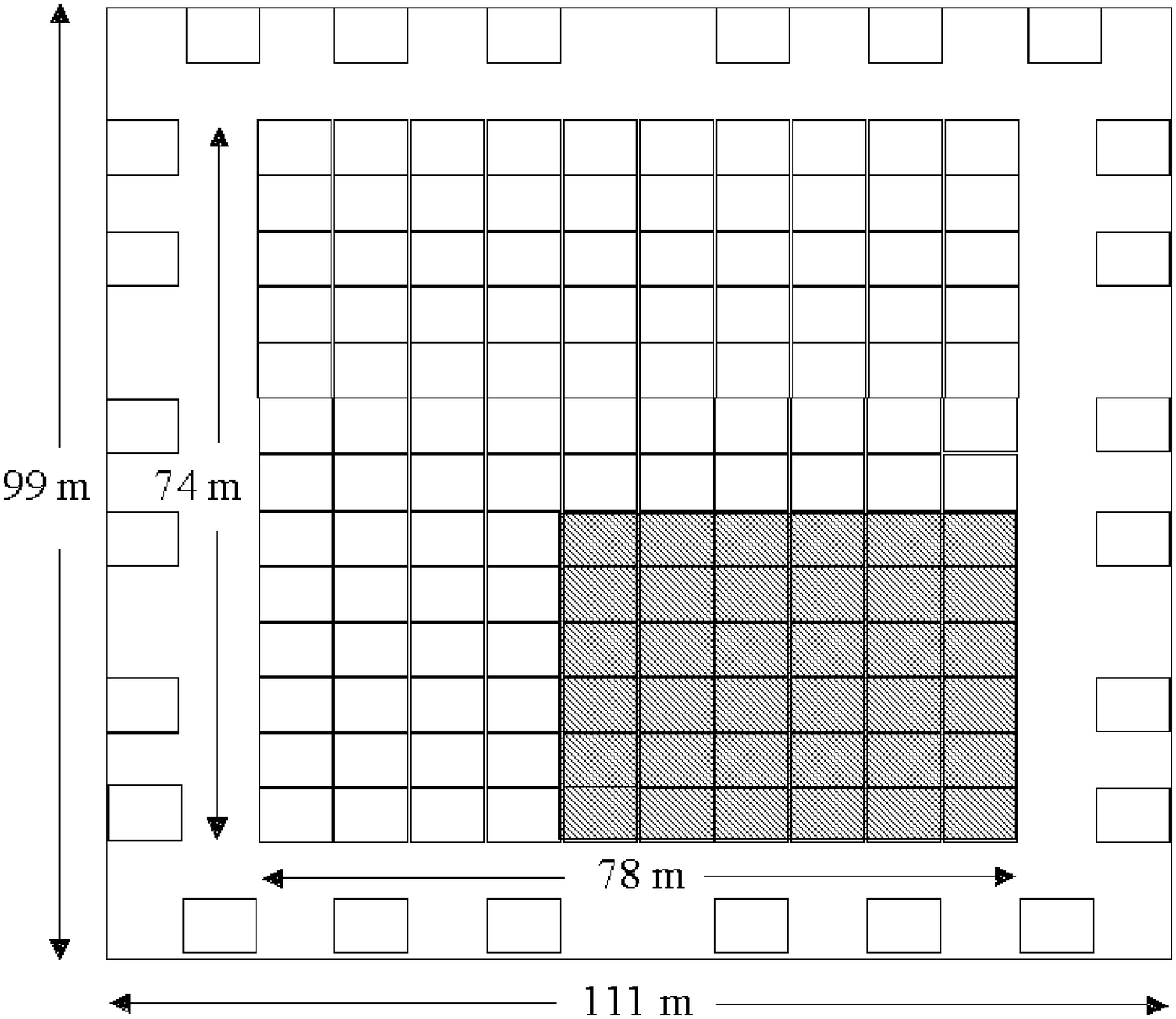}
\end{minipage}
\begin{minipage}[t]{.4\columnwidth}
  \includegraphics[height=.3\textheight]{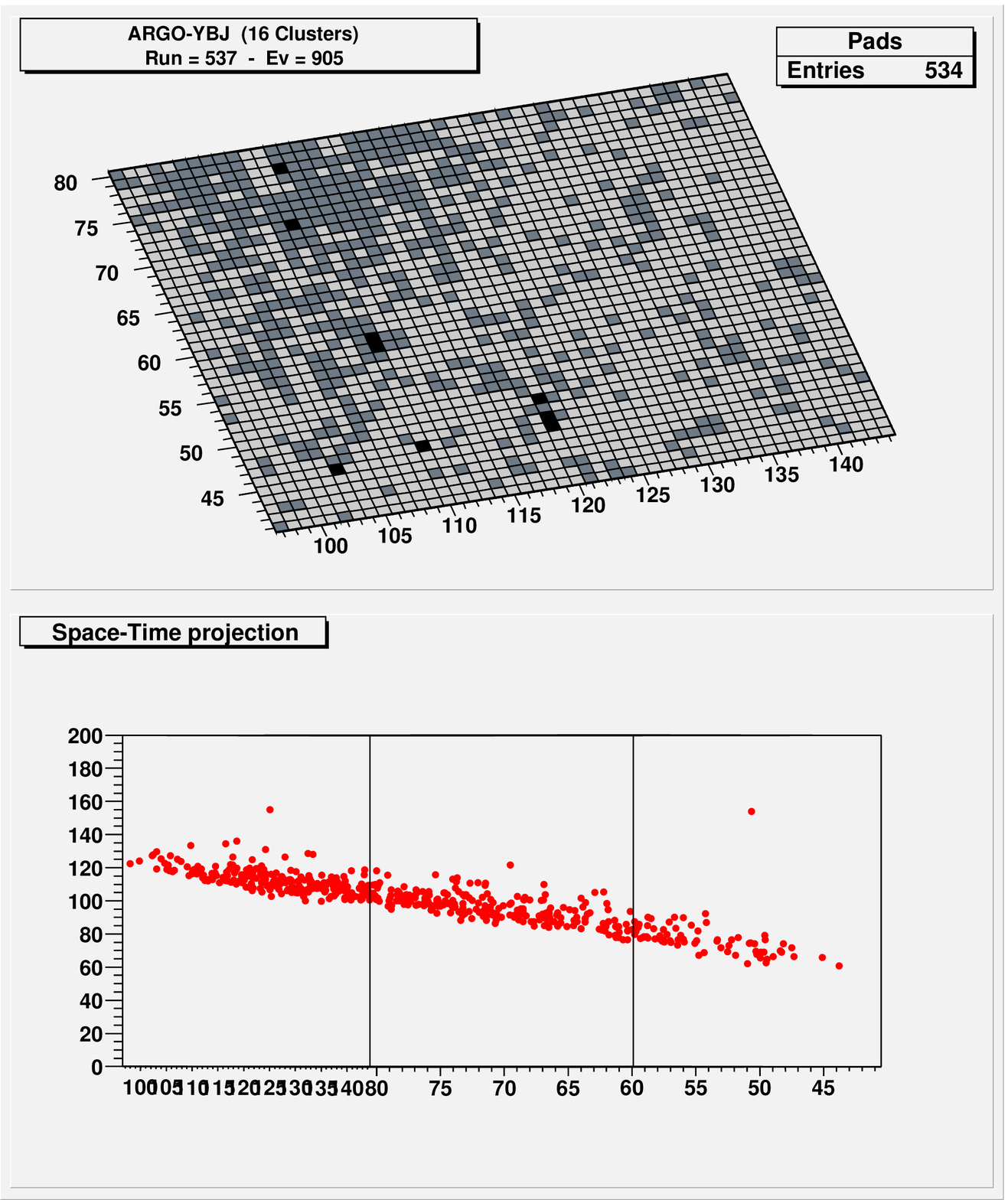}
\end{minipage}
  \caption{Left: Layout of ARGO$-$YBJ, showing the central RPCs carpet and the 
outer ring. The rectangles are subsets of the detector (``clusters'') of area
$\sim 42\;m^2$. The shaded area (36 clusters) is already installed.
Right: Example of a shower recorded with 16 clusters.}
\label{detector}
\end{figure}

The ARGO$-$YBJ detector is currently under construction at the YangBaJing
High Altitude Cosmic Ray Laboratory in Tibet (P.R. of China), 4300 $m$ above
the sea level. It is a full coverage array of dimensions 
$\sim 74\times 78\;m^2$ realized with a single layer of Resistive Plate 
Counters (RPCs). The area surrounding this central detector (``carpet''), up
to $\sim 100\times 110\;m^2$, is partially ($\sim 50\%$) instrumented with
RPCs (``guard ring''), for a total active area of $\sim 6400\; m^2$ 
(see left side of Figure \ref{detector}). The detector basic element is the 
``pad'', of dimensions $56\times 62\;cm^2$, which defines its space-time 
granularity in observing shower fronts. Moreover, the detector is divided in 
$6\times 2$ RPCs units (``clusters'') and covered by a 0.5 $cm$ thick layer 
of lead, in order to convert a fraction of the secondary $\gamma$-rays in 
charged particles, and to reduce the time spread of the shower particles 
\cite{Surdo}. 

\section{Observational techniques}

ARGO$-$YBJ will perform two different types of measurements:

\noindent
a) Shower technique

\noindent
Detection of showers with a trigger threshold $N_{pad} \ge 20$, where 
$N_{pad}$ is the number of fired pads on the detector (``multiplicity''). 
For these events the position and time of any fired pad will be recorded. 
An example of shower with $N_{pad} \sim 500$, recorded with $\sim 10\% $ of the
whole carpet during a test run, is shown in the right side of Figure
\ref{detector}: note the very detailed view of the shower front pattern. 
From the pads data the shower parameters (core position, 
arrival direction and size) can be reconstructed. 

The trigger condition $N_{pad} \ge 20$ corresponds to a primary $\gamma$-ray
energy threshold of a few hundreds GeV, the exact value depending on the
source spectrum and on the zenith angle of the observation. A vertical
$\gamma$-ray of energy $E\sim 120\;GeV$ gives a mean number of particles 
$N_e = 20$ at the ARGO$-$YBJ altitude.

\newpage

\noindent
b) Single particle technique

\noindent
Every $0.5\;s$ the rate of the single particles hitting the detector is
recorded. This measurement allows the detection of the secondary particles
from very low energy showers ($E>10\;GeV$) that reach the ground in a number
insufficient to trigger the detector operating with the shower technique.

\section{High energy emission from GRBs}

So far the only existing data reporting high energy $\gamma$-rays from GRBs
come from the observations of EGRET (however, there are also possible TeV 
emissions claimed in the past by various ground-based experiments, in 
particular that from GRB970417a recorded by the Milagrito detector 
\cite{Milagrito}).
During its lifetime it detected 16 intense events, with a maximum photon
energy of $18\; GeV$ \cite{Catelli}. All their spectra show a power law
behaviour without any cutoff, suggesting that a large fraction of GRBs could
emit GeV or even TeV $\gamma$-rays. 

However, high energy $\gamma$-rays udergo pair production with infrared and
optical stellar photons in the intergalactic space, and are strongly
absorbed during their travel towards the Earth. The optical depth of this
process $\tau (E, z)$ increases with the source redshift $z$ and the
$\gamma$-ray energy $E$. The majority (57\%) of the GRB redshifts measured so 
far (November 2003) is located at $z>1$ and only 2 GRBs have a redshift 
$z<0.2$. According to \cite{Stecker}, at a distance of $z=1(0.1)$ the optical 
depth becomes larger than 1 for $\gamma$-ray energies $E>50(800)\; GeV$. 
Therefore it seems unlikely to detect TeV emission from GRBs and most of the 
efforts must be concentrated in the $10-1000\; GeV$ energy range.

In order to evaluate the ARGO$-$YBJ sensitivity to GRBs we consider a simple
model in which the GRB high energy flux is described by a power law 
spectrum with photon index $\Gamma$ extending up to a maximum energy $E_{max}$
(intrinsic cutoff), and affected by an exponential cutoff due to the 
intergalactic absorption: $dN/dE=KE^{-\Gamma} e^{-\tau (E, z)}$.

The GRB is assumed at a zenith angle $\theta =20^o $ with an intrinsic energy
cutoff in the range $100\; GeV<E_{max} <1\; TeV$ and a distance in the range
$0<z<2$. The absorption factor is calculated exploiting the values of
$\tau (E, z)$ given in \cite{Stecker}. 

The sensitivity has been obtained by comparing the number of events expected
from the GRB with the number of background events, according to both
detection techniques, varying the GRB parameters: spectral normalization
factor $K$, spectrum slope $\Gamma$, cutoff energy $E_{max}$, redshift $z$.

\begin{figure}[t]
  \includegraphics[height=.3\textheight]{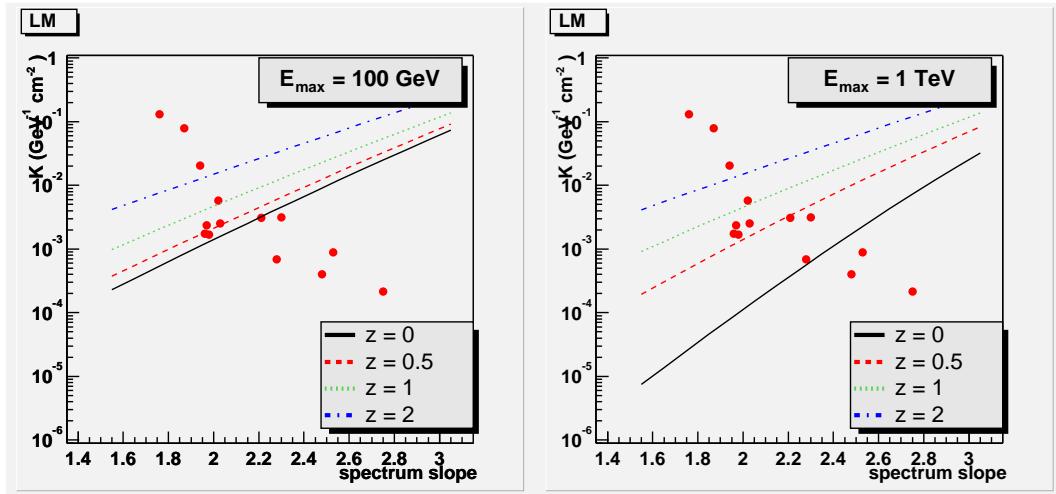}
  \caption{The value of the GRB spectral normalization factor $K$ necessary
to give a $4\sigma$ signal with the LM technique, as a function of the
spectrum slope for different values of $E_{max}$ and $z$. The points
represent 14 GRBs observed by EGRET.}
\label{LM}
\end{figure}

\section{Results}

In the case of observation technique a), a GRB candidate will appear as a 
statistical significant excess of Low Multiplicity (LM) events clustered in
time and arrival direction. The angular resolution of the detector 
({\it i.e.}, the opening angle containing 71.5\% of the $\gamma$-ray events)
for showers with $N_{pad} \ge 20$ is $\sim 2.7^o $.

In the case of observation technique b), no reconstruction of the shower
parameters is possible and a GRB is observed as an excess in the Single
Particle (SP) background rate, possibly in time coincidence with a GRB
satellite detection \cite{Silvia}.

\begin{figure}
  \includegraphics[height=.3\textheight]{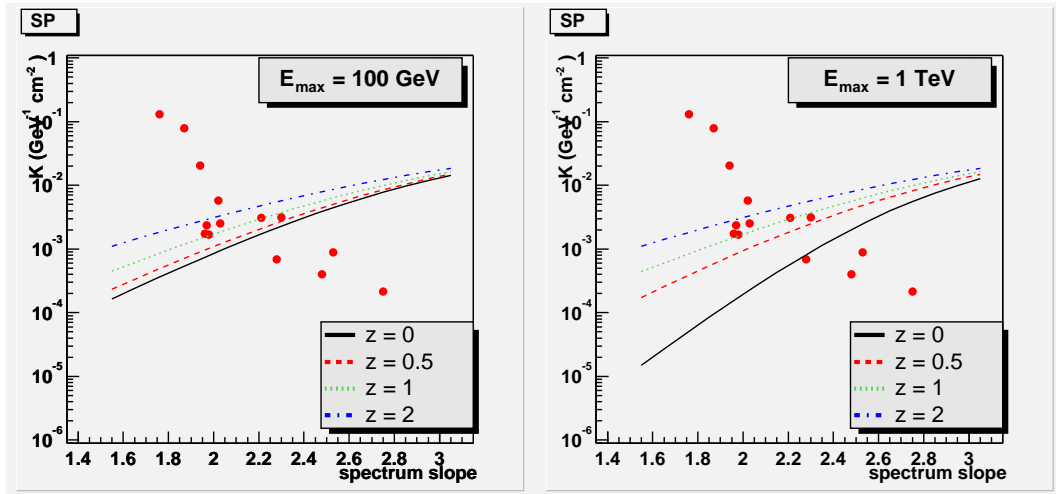}
  \caption{Same as Figure \ref{LM} but in the case of the SP technique.}
\label{SP}
\end{figure}

Our results are summarized in Figure \ref{LM} (for the LM technique) and
in Figure \ref{SP} (for the SP technique), where the value of $K$ necessary
to give a signal with a statistical significance of $4\sigma$ is shown as a
function of the spectrum slope $\Gamma$ for different values of $E_{max}$
and $z$. In these calculations a GRB duration $\Delta t=1\; s$ is assumed.
The sensitivity for different durations can be easily obtained by
multiplying $K$ by $\sqrt{\Delta t}$. To compare the ARGO$-$YBJ expected
sensitivity with real GRBs, the $K$ vs. $\Gamma$ values of 14 EGRET
GRBs \cite{Catelli} are plotted in the same figures.

These results show that the SP technique is in general more sensitive to GRBs, 
in particular for high $z$ where the intergalactic absorption strongly affects 
the high energy tail of the spectrum. Only in the case $z=0$ and 
$E_{max} =1\; TeV$ the LM shower technique is slightly better.

\section{Conclusions}

ARGO$-$YBJ could observe the high energy emission of the most intense
GRBs. Since $\gamma$-rays of energy $E>1\; TeV$ emitted by cosmological
sources are strongly absorbed during their travel towards the Earth, the
SP technique provides the best approach to detect GRBs, being particularly
sensitive in the $10-1000\; GeV$ energy range. 

The sensitivity and the event rate depend critically on the shape of the
GRBs spectra above 10 GeV. This shape is determined by the possible existence
of an intrinsic cutoff and by the absorption of $\gamma$-rays in the
intergalactic space. 

The analysis of 14 EGRET GRBs indicates that if the intrinsic cutoff is not
too low ($E_{max} >100\; GeV$) and the sources redshift is $z<2$, a 
fraction of the events ranging from $\sim 20\%$ to $\sim 80\%$
would be detectable.

\end{document}